\documentclass[aps,prl,twocolumn,showpacs, preprintnumbers,amsmath,amssymb, superscriptaddress]{revtex4}  

\usepackage{latexsym}
\usepackage{graphics}
\usepackage{dcolumn}
\usepackage{bm}

\begin{document}

\preprint{APS/123-QED}

\title{Direct observations of nucleation in a nondilute multicomponent alloy}

\author{Chantal K. Sudbrack}
\email{csudbrack@alumni.reed.edu}
\affiliation{
Northwestern University, Department of Materials Science and Engineering, Evanston, IL 60208  USA}

\author{Ronald D. Noebe}
\affiliation{NASA Glenn Research Center, Cleveland, OH 44135  USA}

\author{David N. Seidman}
\email{d-seidman@northwestern.edu}
\affiliation{
Northwestern University, Department of Materials Science and Engineering, Evanston, IL 60208  USA}

\date{\today}

\begin{abstract}

The chemical pathways leading to $\gamma$'(L1$_2$)-nucleation from \textit{nondilute} Ni-5.2 Al-14.2 Cr at.\%, $\gamma$(f.c.c.), at 873 K are followed with radial distribution functions and isoconcentration surface analyses of direct-space atom-probe tomographic images. Although Cr atoms initially are randomly distributed, a distribution of congruent Ni$_3$Al short-range order domains (SRO), $<$\textit{R}$>$$\cong$0.6 nm, results from Al diffusion during quenching. Domain site occupancy develops as their number density increases leading to Al-rich phase separation by $\gamma$'-nucleation, $<$\textit{R}$>$=0.75 nm, after SRO occurs.
\end{abstract}

\maketitle


A fundamental aspect of any phase transition involves the nucleation of stable nuclei. 
The theoretical nucleation literature is vast, e.g. \cite{wagner01, russell80, soisson00}, whereas direct experimental observations of sub-nano to nanoscale nuclei are miniscule \cite{aaronson92}, controversial \cite{stowell02}, and limited to binary systems \cite{aaronson92}. Experimental studies to date have measured the kinetics of nuclei with radii, \textit{R}, $>$ 2 nm \cite{aaronson92}, with no knowledge of the chemical pathways associated with nucleation.  With atom-probe tomography (APT), it is possible to characterize \textit{both} the spatial extent and compositions of nuclei as small as 0.45 nm \cite{sudbrack04} with direct-space atomic reconstructions of small volumes of material  (typically 2x10$^4$ nm$^3$ containing 10$^6$ atoms). We demonstrate herein that APT can also follow the chemical pathways prior to (i.e. ordering and clustering) and during nucleation of Ni$_3$(Al$_x$Cr$_{1-x}$) $\gamma$'-phase (L1$_2$) in a moderately supersaturated $\gamma$-matrix (f.c.c.) with isoconcentration surfaces \cite{hellman03} and  radial distribution function (RDF) \cite{marquis03} analyses.

\begin{figure}
\includegraphics{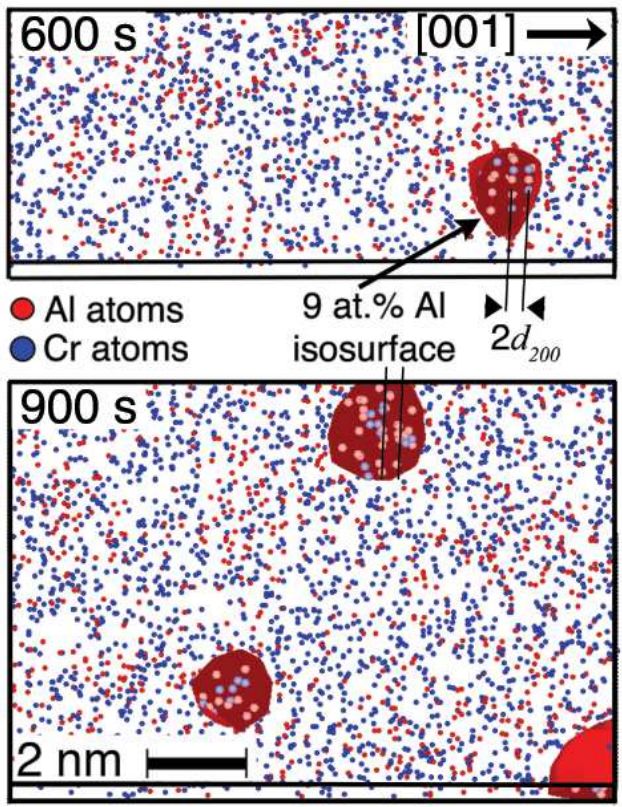}
\caption{\label{fig1}Within Ni-5.2 Al-14.2 Cr at.\% APT reconstructed volumes, Al-rich $\gamma$'-precipitates are first detected with 9 at.\% Al isoconcentration surfaces after aging for 600 s (\textit{top}) at 873 K.  Between 600 s (\textit{top}) and 900 s (\textit{bottom}), the $<$\textit{R}$>$ remains constant (0.75 nm, $\sim$150 atoms); however, a sharp increase in number density from 3.6x10$^{23}$ m$^{-3}$ to 2.1x10$^{24}$ m$^{-3}$ is observed . The Al and Cr atoms in these 3 nm thick \textit{partial} volumes, displayed in red and blue, are enlarged in the $\gamma$'-precipitates to emphasize the resolved \{002\} superlattice planes perpendicular to the [001] analysis direction, where the average $\gamma$'-composition is Ni-18.3$\pm$0.9 Al-9.3$\pm$0.7 Cr at.\%.}
\end{figure}


High purity elements were induction melted in Ar and then chill cast in a copper mold (diam.=19 mm). Cast ingots are homogenized in Ar at 1573 K for 24 h and then 1123 K for 3 h; subsequently 10 mm thick water-quenched sections are aged at 873 K (120 s to 1024 h). Aged specimens are electropolished to produce sharply-pointed needle-shaped specimens for APT  \cite{sudbrack04}. The solute distributions in 3 nm thick \textit{partial} APT volumes (Fig.~\ref{fig1}) provide direct visual evidence of phase separation (i.e. $\gamma$'-nucleation) and demonstrates that atoms are positioned with sub-nanometer resolution with APT.  Employing 9 at.\% Al isoconcentration surface analyses, spheroidal Al-rich regions with \textit{R} as small as 0.45 nm, 20 \textit{detected} atoms (ion detection efficiency of 60\%), are first observed within APT images at 600 s. Along the [001]-direction in L1$_2$ Ni$_3$(Al$_x$Cr$_{1-x}$), nearly pure Ni-rich planes alternate with solute-rich planes. This alternation is captured within the delineated Al-rich regions, i.e. the Al-rich \{002\}-superlattice planes (\textit{2d}$_{200}$=0.356 nm) are resolved, thus providing further evidence that these regions are $\gamma$'-precipitates. Between 600 and 900 s,  $\gamma$'-number density, \textit{N$_v$}, \textit{increases} sharply from 3.6x10$^{23}$ to 2.1x10$^{24}$ m$^{-3}$ at a constant $<$\textit{R}$>$ of 0.75 nm, implying that the transformation is undergoing quasi-steady-state nucleation. Their average composition, Ni-18.3$\pm$0.9 Al-9.3$\pm$0.7 Cr at.\%, estimates a lower-bound for the $\gamma$'-critical nucleus. We find the $\chi$$^2$-test \cite{pareige99} is inadequate to evaluate non-randomness of the Al and Cr distributions obtained with APT.  Alternatively, RDF analyses offer a higher degree of spatial sensitivity and resolve lattice site-specific information, thereby allowing the chemical pathways associated with $\gamma$'-nucleation (\textit{t} $\leq$ 600 s) to be followed in direct-space for the first-time.

An RDF at a given radial distance, \textit{r}, is defined as the average concentration distribution of component \textit{i} around a given solute species, X,  $<$\textit{C$_i^X$}(\textit{r})$>$, normalized to the overall concentration of \textit{i} atoms, \textit{C$_i^o$}, in the sampled volume: 
\begin{equation}
RDF=
\frac{<\textit{C$_i^X$}(\textit{r})>}{\textit{C$_i^o$}} =
\frac{1}{\textit{C$_i^o$}}
\sum_{k=1}^{\textit{N$_X$}}
\frac{\textit{N$_i^k$}(\textit{r})}{\textit{N$_{TOT}^k$}(\textit{r})}; 
\label{eq1}
\end{equation}	
where \textit{N$_i^k$}(\textit{r}) is the number of \textit{i} atoms in a radial shell around the \textit{k}$^{th}$ X atom that is centered at \textit{r}, \textit{N$_{TOT}^k$}(\textit{r}) is the total number of atoms in this shell, and \textit{N$_X$} is the number of  X atoms in this volume. The average concentration distributions around a solute species in 0.01 nm thick shells are smoothed with a weighted moving average using a Gaussian-like spline-function \cite{hellman00}, with a 0.04 nm full-width half-maximum, where $<$\textit{N$_{TOT}$}(\textit{r})$>$ increases quadratically with \textit{r} and the total volume sampled for a given shell is proportional to X.  Only the RDFs for \textit{r} $\geq$ 0.2 nm are presented as the physical interpretation at smaller \textit{r} is difficult due to \textit{possible} ion trajectory effects. RDF values of unity describe perfectly random distributions, while values that differ describe clustering or ordering. The absolute magnitude of these processes can be compared with the RDF's amplitude, \textit{A}=[RDF(\textit{r})--1]; where \textit{A} $>$ 0 indicates a greater concentration than the overall concentration (positive correlation) and \textit{A} $<$ 0 implies a smaller one (negative correlation).

\begin{figure}
\includegraphics{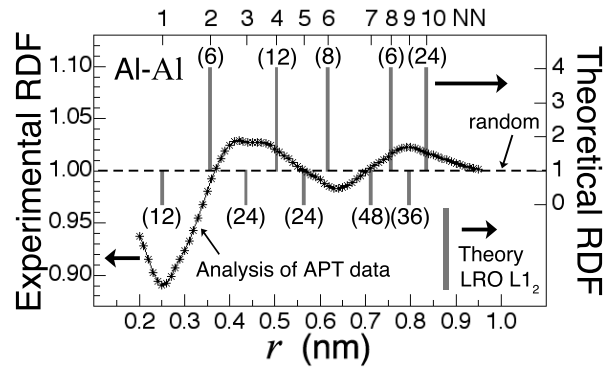}
\caption{\label{fig2}Al-\textit{Al} RDFs (Eq.~\ref{eq1}) vs. radial distance, \textit{r}, out to the 10$^{th}$ nearest neighbor (NN) distance for a nearly stoichiometric Ni$_3$Al (L1$_2$) alloy \cite{apfimnote} analyzed by APT and a theoretical alloy assuming that the LRO=1.  The number of L1$_2$ NNs at corresponding distances is indicated in parentheses. RDF= 1 (random solid solution) is indicated by a horizontal dashed line.}  
\end{figure}

To evaluate an RDF's shape and magnitude when analyzing APT data, which is affected by the APT's positional uncertainty, we first consider a well-defined system; a nearly-pure Ni$_3$Al (L1$_2$) alloy \cite{apfimnote} that exhibits almost perfect long-range order (LRO) \cite{vanbakel95} and compare it to the expected values for a theoretical Ni$_3$Al alloy with LRO=1, Fig.~\ref{fig2}. In the theoretical alloy, the Al-\textit{Al} profile (Eq.~\ref{eq1}) alternates discretely as a series of delta-functions between \textit{A}=-1 and \textit{A}=3 at successive nearest neighbor (NN) distances (grey vertical lines in Fig.~\ref{fig2}), while the Al-\textit{Ni} profile alternates between 1/3 and -1 (not displayed).  Because mass conservation obtains, one RDF in a binary alloy is sufficient to describe the LRO behavior; therefore, only the Al-\textit{Al} profiles are presented. The random-state, RDF=1, is denoted by a dashed horizontal line. Compared to the theoretical profile (\textit{right-hand} ordinate), the experimental profile (\textit{left-hand} ordinate) is damped (ca. nine times smaller in \textit{A} than the theoretical value at the 1$^{st}$ NN distance) and does not follow exactly the theoretical positive and negative alternation. The experimental values of the Al-\textit{Al} and  Al-\textit{Ni} RDFs are strongest at the 1$^{st}$ NN distances than at larger \textit{r}. These features of the RDF result from contributions of adjacent NNs to the measured concentrations and is a function of the number of NNs (parenthetical values in Fig.~\ref{fig2}) and their weighted contribution (proportional to theoretical \textit{A} values). Note the measured RDFs oscillate over the full range of \textit{r}, \textit{r} $\leq$ 1 nm, which is a distinguishing feature of LRO in an APT RDF. The differences with theory are understood by considering the value for a particular NN.  As the RDF in  Eq.~\ref{eq1} is not normalized to the coordination number, \textit{Z} (as \textit{Z} assignment is not straightforward due to positional uncertainty), a greater number of NNs contribute more to the measured RDF than fewer with the same weight. At the 3$^{rd}$ NN distance, although theory predicts a negative correlation, the observed positive correlation results from the positive contributions from the six adjacent 2$^{nd}$ NNs and 12 4$^{th}$ NNs, each with a larger weighting factor (\textit{A}=3) than the 24  3$^{rd}$ NNs  (\textit{A}=-1). Nevertheless the experimental results in Fig.~\ref{fig2} are indicative of LRO and act as a reference state for what follows. If the distributions of reconstructed atom positions about each NN distance are known, it would be possible to obtain a deconvoluted RDF, allowing pair-wise interaction energies to be determined experimentally in multicomponent systems.

\begin{figure}
\includegraphics{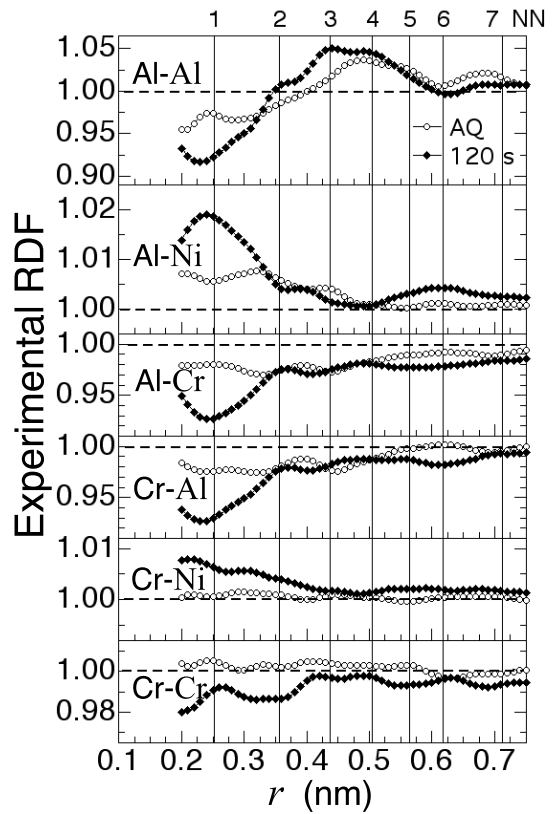}
\caption{\label{fig3}Experimental RDF vs. radial distance, \textit{r}, out to the 7$^{th}$ NN distance for as-quenched (AQ) Ni-5.2 Al-14.2 Cr at.\% and after aging for 120 s at 873 K.  The negative and positive correlations at the 1$^{st}$ NN in the Al-\textit{Al} and Al-\textit{Ni} profiles establish the presence of L1$_2$-order.}
\end{figure}

Figure~\ref{fig3} presents experimentally determined Al- and Cr-centered RDFs for the as-quenched (AQ) state (open circles) and after 120 s (solid diamonds) of aging. The comparison to the random state (dashed horizontal line) checks the homogeneity of the post-quenching state. The nearly flat Cr-\textit{Cr} and Cr-\textit{Ni} AQ profiles indicate that the quench-rate is sufficient to produce a \textit{random} Cr distribution relative to other Cr and Ni atoms. The initial quench rate is, however, insufficient to avoid diffusion of Al away from Cr, as Al diffuses considerably faster than Cr does in Ni-Al-Cr. The Al-\textit{Al} and Al-\textit{Ni} AQ RDFs' deviations from the random state are comparable to the ones for the LRO alloy, Fig.~\ref{fig2}. Unlike the LRO alloy they occur over a shorter range, \textit{r} $\leq$ 0.6 nm. At the 1$^{st}$ NN distance, the Al-\textit{Al}  and Al-\textit{Ni} correlations are negative and positive, respectively, and the Al-\textit{Al} RDF's oscillations are coupled with oscillations in the Al-\textit{Ni} RDF that are opposite in sign, which is consistent with L1$_2$ ordering. These trends establish the presence of quenched-in L1$_2$ SRO domains with $<$\textit{R}$>$$\cong$0.6 nm. After 120 s, an increase in the Al-\textit{Al} and Al-\textit{Ni} RDFs' intensities, which also extends 0.6 nm, indicates the SRO in solution increases. It follows that the SRO domains' $<$\textit{R}$>$ remains ca. constant as their \textit{N$_v$} increases concomitantly. Since the solute concentrations are nondilute, the volumes are reasonably well-sampled, yielding Al-\textit{Cr} and Cr-\textit{Al} profiles that are very similar. These RDFs exhibit a negative spatial correlation over the distances analyzed, \textit{r} $<$ 1 nm, thereby establishing that the $\gamma$'-SRO regions are Cr depleted. In concert with more developed SRO at 120 s, the intensity of the Al-\textit{Cr} RDF at the 1$^{st}$ NN distance is strongly negative, which suggests that Cr atoms do \textit{not} prefer to be NNs to Al and that they occupy the Al sites within the Ni$_3$(Al$_x$Cr$_{1-x}$) SRO domains. A \textit{slight} Ni-Cr ordering tendency, either within Al-rich domains or in the $\gamma$-matrix, is observed at 120 s.   

\begin{figure}
\includegraphics{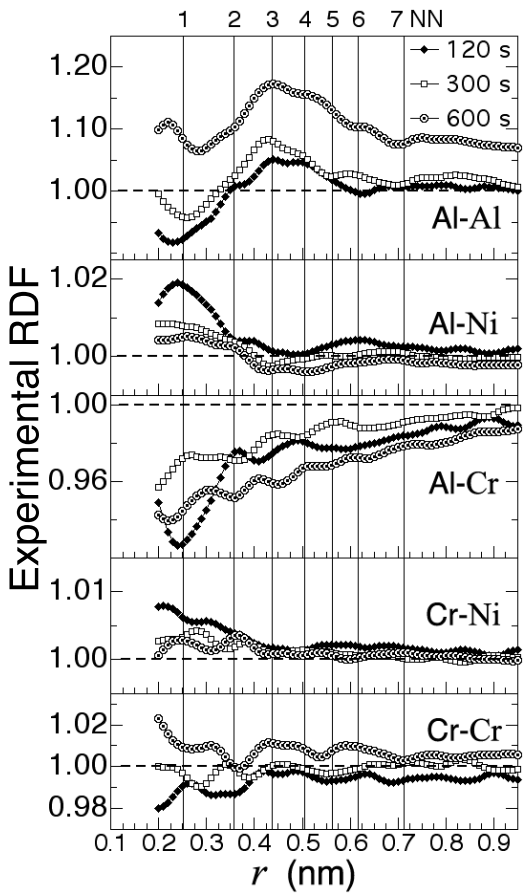}
\caption{\label{fig4}Experimental RDF vs. radial distance, \textit{r}, for Ni-5.2 Al-14.2 Cr at.\% aged for 120, 300, or 600 s at 873 K. The overall Al-\textit{Al} shift to higher RDF values after 300 s corresponds to the initiation of phase separation by $\gamma$'-nucleation.}
\end{figure}

Since the RDF is applied to total volumes, as the alloy decomposes (\textit{t}$\geq$120 s, Fig.~\ref{fig4}) the RDF becomes a convolution of the ordering and clustering processes in both $\gamma$- and $\gamma$'-phases, where cumulative shifts in the RDF away from unity (Eq.~\ref{eq1}) correspond to elemental partitioning associated with $\gamma$'-nucleation. Isoconcentration surfaces (Fig.~\ref{fig1}) first detect a small volume fraction (0.11$\pm$0.04 \%) of $\gamma$'-nuclei (\textit{R}$\geq$0.45 nm) that are highly enriched in Al (19.1$\pm$1.7 \%) at 600 s. Correspondingly the Al-\textit{Al} RDF shifts strongly. In comparison, the shift at 300 s is slight (no appreciable phase separation) and no $\gamma$'-nuclei are detected with isoconcentration surfaces, establishing that the onset of $\gamma$'-phase separation occurs between 300 and 600 s. For \textit{t}=0 (Fig.~\ref{fig3}), 120 and 300 s (Fig.~\ref{fig4}), since the Al-\textit{Al} profiles oscillate about unity, the SRO domains observed \textit{prior to phase separation} are congruently ordered, i.e., their compositions do not deviate from the overall concentration. The strong anti-site Cr preference for Al in the domains at 120 s diminishes with further aging, where a decrease in negative correlation at the 1$^{st}$ NN distance in the Al-\textit{Cr} RDFs demonstrates that Al substitutes for Cr in the SRO domains leading to less retained Cr. Unlike for Al-\textit{Al} RDFs, the shift in Cr-\textit{Cr} RDFs with nucleation is gradual and is positive as Cr partitions to $\gamma$, where a slight negative correlation at 120 s may correspond to Cr ordering in the domains.  A strong Al-\textit{Ni} ordering tendency at the 1$^{st}$ NN distance at 120 s damps toward unity with aging, establishing that the cumulative L1$_2$-ordering (SRO and LRO) in $\gamma$ + $\gamma$' diminishes. Since $\gamma$'-nucleation clearly leads to localized LRO development, the observed Al-\textit{Ni} RDF damping corresponds to \textit{SRO diminishment in the $\gamma$-matrix with aging}.

Lattice kinetic Monte Carlo (LKMC) simulations \cite{pareige99} of a nearly identical alloy, Ni-5.2 Al-14.8 Cr at.\%, aged at 873 K, indicate that $\gamma$'-phase separation occurs by nucleation and growth, which is preceded by Ni$_3$Cr-type (DO$_{23}$) SRO that develops, decays, and is followed by  L1$_2$ SRO development.  We \textit{do not} observe this behavior experimentally, but find that the RDFs establish L1$_2$ SRO in the AQ state, which becomes stronger after 120 s, Fig.~\ref{fig3}. In the AQ alloy, the distribution of Cr and Ni atoms relative to Cr atoms is initially random, hence Ni$_3$Cr SRO \textit{does not} precede the L1$_2$ SRO.  The onset of Al-rich $\gamma$'-phase separation (Fig.~\ref{fig4}) occurs \textit{after} 300 s of aging after SRO has occurred. A crucial difference between our experiments and the LKMC simulation \cite{pareige99} is the thermal history of the AQ state. In the LKMC simulation the initial state is a random distribution of atoms, which neglects the influence of a finite quench rate from 1123 K, followed by up-quenching to 873 K. The influence of the quench rate is twofold. Firstly, it leads to an unknown value for the supersaturation of quenched-in vacancies. Secondly, phase separation may occur, particularly in alloys that are highly supersaturated and/or have a small interfacial free-energy, where the nucleation barriers are small and nucleation currents large \cite{wagner01, russell80, soisson00}. For dislocation densities ranging between 10$^{10}$ to 10$^{12}$ m$^{-2}$, the half-life time for the annihilation of excess vacancies at 873 K, for diffusion-limited annihilation, is between 10118 and 107 s, which are lower bounds since dislocation climb is assumed to be diffusion-limited \cite{seidman66}, whereas it need not be \cite{seidman67}. This suggests that quenched-in excess vacancies may play a role in the early-stages of decomposition.

In summary, the chemical pathways leading to $\gamma$'-nucleation at 873 K from a solutionized nondilute Ni-Al-Cr alloy are characterized in detail with isoconcentration surfaces and RDF (Eq.~\ref{eq1}) analyses of direct-space APT images, Figs. 1--4. Our new approach offers the unique ability to characterize these pathways in nondilute multicomponent alloys, which are inherently complex, in particular when elemental diffusivities differ significantly. The application of an RDF to APT images provides a highly sensitive methodology to test homogeneity, which despite the APT's uncertainty in positioning absolutely atoms allows lattice-site ordering and clustering to be investigated, Fig.~\ref{fig2}. An alloy's thermal history, in addition to quenched-in vacancies, influences the chemical pathways, which differ from predictions by LKMC simulations assuming a completely random solid-solution \cite{pareige99}. The finite quench-rate after solutionizing is sufficient to yield a random distribution of Cr atoms, Fig.~\ref{fig3}. Quenching, however, produces a distribution of Ni$_3$(Al$_x$Cr$_{1-x}$) SRO (L1$_2$) domains ($<$\textit{R}$>$$\cong$0.6 nm) that are congruently-ordered and Cr-depleted relative to the solid solution. Initially, trapped Cr atoms show an anti-site preference for the Al sublattice. With aging, as the \textit{N$_v$} of SRO domains increases and then $\gamma$'-nuclei form, Cr is expelled and substituted for by Al. As highly localized LRO regions are established ($\gamma$'-nuclei), the SRO in the $\gamma$-matrix diminishes. A strong shift in Al-\textit{Al} RDF (Fig.~\ref{fig4}) coincides with the detection of $\gamma$'-nuclei (Fig.~\ref{fig1}), thereby establishing that the onset of phase separation occurs between 300 and 600 s after SRO establishment. Between 600 and 900 s, nuclei as small as \textit{R}=0.45 nm are detected, while the \textit{N$_v$} of $\gamma$'-nuclei increases sharply as their  $<$\textit{R}$>$ value remains constant (0.75 nm), Fig.~\ref{fig1}. 

This research is sponsored by the National Science Foundation, grant DMR-0241928; CKS received partial support from an NSF graduate fellowship. We thank Dr. K. E. Yoon for providing some APT data, and Dr. Z. Mao and Prof. M. Asta for helpful discussions.


\end{document}